\documentclass[twocolumn,prl,tightenlines,superscriptaddress]{revtex4-2}
\usepackage{hyperref}
\usepackage{amsmath,amssymb}
\usepackage[utf8]{inputenc} 			
\usepackage[T1]{fontenc}				
\usepackage{graphicx}
\usepackage{float}
\usepackage[usenames,dvipsnames]{xcolor}
\usepackage{tikz}
\usepackage{ulem}

\renewcommand{\div}{\nabla\cdot}
\renewcommand{\vec}[1]{\mathbf{#1}}

\newcommand*\colvec[3][]{
    \begin{pmatrix}\ifx\relax#1\relax\else#1\\\fi#2\\#3\end{pmatrix}
}

\newcommand{\nperp}{\vec n_\bot}
\newcommand{\dpar}{\partial_\parallel}
\newcommand{\dperp}{\partial_\bot}

\begin{document}

\title{Continuing the discussion on ``The inconvenient truth about flocks'' by Chen {\it et al.}}

\author{Hugues Chat\'{e}}
\affiliation{Service de Physique de l'Etat Condens\'e, CEA, CNRS Universit\'e Paris-Saclay, CEA-Saclay, 91191 Gif-sur-Yvette, France}
\affiliation{Computational Science Research Center, Beijing 100094, China}
\affiliation{Sorbonne Universit\'e, CNRS, Laboratoire de Physique Th\'eorique de la Mati\`ere Condens\'ee, 75005 Paris, France}

\author{Alexandre Solon}
\affiliation{Sorbonne Universit\'e, CNRS, Laboratoire de Physique Th\'eorique de la Mati\`ere Condens\'ee, 75005 Paris, France}

\date{\today}

\begin{abstract}
  We hereby reply concisely and hopefully clearly to the ongoing
  claims of incorrectness made by Chen et al. about our work on
  two-dimensional flocks.
\end{abstract}

\maketitle

In~\cite{chate2025comment}, we summarized our disagreements with ``The
inconvenient truth about flocks'' by Chen et
al.~\cite{maitra_inconvenient_2025}, especially regarding our earlier
publication on the hydrodynamic description of polar ordered
flocks~\cite{chate_dynamic_2024}. We find that the recent reply by
Chen at al~\cite{chen2025response} to our comments shows that our work
has been misunderstood. We thus take the opportunity here to,
hopefully, clarify this inconvenient situation, and refute some claims
of~\cite{chen2025response}.

{\it Continuity equation.} For the sake of completeness, let us first repeat the argument
of~\cite{chate_dynamic_2024,chate2025comment} showing that, for two-dimensional
systems spontaneously breaking the rotational symmetry, the
deterministic part of the dynamics of the Goldstone mode
$\theta(\vec r,t)$ must be written as a continuity equation
\begin{equation}
  \label{eq:NG-theta}
  \partial_t \theta(\vec r,t)=-\div \vec J+\eta(\vec r,t)
\end{equation}
with $\eta$ a Gaussian white noise, delta correlated in space.

Eq.~(\ref{eq:NG-theta}) is most simply argued for by considering the
global direction $\Theta(t)=\frac{1}{V}\int \theta(\vec r,t) d\vec r$
where the integration is over the full volume $V$ of the system. For a
spontaneous symmetry breaking, the dynamics of $\Theta(t)$ must be
pure noise $\partial_t\Theta(t)=\eta_V(t)$. Indeed, any deterministic
relaxation of $\Theta(t)$ toward a different direction would directly
violate the fact that it is the only direction picked by symmetry
breaking. The noise $\eta_V$ can be assumed Gaussian since it is the
sum of the local noise at every point in space, and white if one
considers a large enough coarse-graining scale that all finite time
scales have been integrated out. Repeating the same argument for a
finite part of the system $\mathcal{V}$ delimited by surface
  $\mathcal{S}$, there is an additional term coming from the
interaction with the rest of the system which, if interactions are
local, takes the form of a surface integral
$\partial_t\int_{\mathcal V}\theta(\vec
r,t)=-\int_{\mathcal{S}}\vec n\cdot \vec J+\eta_{\mathcal{V}}(t)$. In
local form, this is exactly Eq.~(\ref{eq:NG-theta}).

The argument above has obvious limitations. It does not apply to the
rotating steady states of chiral systems~\cite{daviet2024kardar},
which do not break rotational symmetry, or to systems where rotational
symmetry is explicitly broken, for example if translational symmetry
is broken~\footnote{As is well known~\cite{chaikin1995principles},
  there is then no additional Goldstone mode associated with
  rotational symmetry breaking.} or if an external field or gradient
is imposed.  This includes solids, smectics, discotics, clock models,
time-crystalline order and systems with an externally fixed density
gradient, all wrongly invoked as counterexamples to
Eq.~(\ref{eq:NG-theta}) by Chen et
al~\cite{maitra_inconvenient_2025,chen2025response}.

Note that, as clear from the argument above, the fact that the noise
is not conserved in Eq.~(\ref{eq:NG-theta}) is required for the
symmetry class that we consider, in which the global direction is not
a conserved quantity. It is certainly not ``inconsistent'', as bluntly
asserted in~\cite{chen2025response}.

{\it Hydrodynamic equation.} A large part of the disagreement with
Chen et al. comes from the fact that the Goldstone mode is a
collective mode, and Eq.~(\ref{eq:NG-theta}) should be thought of as a
coarse-grained equation, obtained after integrating out the fast
degrees of freedom. (Formally, it is the fixed point of the
renormalization group (RG) flow.) Hence, it shall not surprise us that
continuum equations which describe the system on smaller length
scales, possibly with additional fast degrees of freedom, cannot be
written as a continuity equation like Eq.~(\ref{eq:NG-theta}). This is
the case, for example, of Eqs.(IV.5-6) and (IV.17-18)
in~\cite{maitra_inconvenient_2025} which contain an infinite number of
non-linearities, most of which are irrelevant and thus disappear from
the large-scale ``hydrodynamic'' description.

This confusion between the hydrodynamic and continuum description of a
system also underlies our disagreement
regarding~\cite{martin_unified_1972}. There, Martin, Parodi and
Pershan write the (linear) hydrodynamic description of systems in
different symmetry classes, which contains the relevant terms and
associated transport coefficients, and is in agreement with
Eq.~(\ref{eq:NG-theta}) for the classes where it applies. The claim of
Chen et al.~\cite{chen2025response} (but not of Martin, Parodi and
Pershan) is that, at the nonlinear level, the transport coefficients
should be promoted to functions of the fields, which would then
violate Eq.~(\ref{eq:NG-theta}). We respectfully disagree with this
proposal. Even at nonlinear level, the hydrodynamic description should
contain a small number of relevant terms with associated constant
coefficients and not arbitrary dependencies in the fields.

For some systems, the coarse-graining from a small-scale continuum
equation to the collective dynamics Eq.~(\ref{eq:NG-theta}) has been
done explicitly using RG. These directly illustrate how the continuity
Eq.~(\ref{eq:NG-theta}) emerges from an equation which does not obey
this symmetry.

A first example is provided by the work of Nelson and
Pelcovits~\cite{nelson_momentum-shell_1977} on two-dimensional nematic
liquid crystals. They start from the Frank free energy which reads in
2d
\begin{equation}
  \label{eq:frank}
  F[\vec n]=\int d\vec r \left(\frac{K_1}{2}\left[\div\vec n\right]^2+\frac{K_3}{2}\left[\vec n\times(\nabla\times\vec n)\right]^2\right)
\end{equation}
for a nematic director field $\vec n(\vec r)$. Parameterizing by an
angle $\theta$, $\vec n=(\cos\theta,\sin\theta)$, one can write the
associated relaxational (model A) dynamics
\begin{align}
  \label{eq:nematic-modelA}
  \partial_t&\theta(\vec r,t)=-\frac{\delta F}{\delta \theta(\vec r,t)}+\eta(\vec r,t)\\
                            &=(K_1-K_3)\dpar\theta\dperp\theta+\div\left( K_1\dperp\theta \nperp+K_3\dpar\theta\vec n\right)+\eta\nonumber
\end{align}
where $\eta$ is a Gaussian white noise with correlations
$\langle \eta(\vec r,t)\eta(\vec r',t')\rangle=2 T\delta(\vec r-\vec
r')\delta(t-t')$ with temperature $T$, and we have used the notation
$\dpar\equiv \vec n\cdot\nabla$ and $\dperp\equiv \nperp\cdot\nabla$
with
$\nperp\equiv(-\sin\theta,\cos\theta)$. Eq.~(\ref{eq:nematic-modelA})
is not a continuity equation because of the term proportional to
$K1-K_3$ but, as shown in~\cite{nelson_momentum-shell_1977}, the RG
flow goes towards a fixed point with $K_1=K_3$ so that the symmetry
Eq.~(\ref{eq:NG-theta}) is realized on large scales, consistently with
our claim. That Ref.~\cite{nelson_momentum-shell_1977} is used by Chen
et al.~\cite{chen2025response} against our theory only shows that it
was misunderstood.

A second example is found in the work of Jentsch and Lee about the
homogeneous flocking phase~\cite{jentsch_new_2024}. At the price of
neglecting several non-linearities to keep the problem tractable, they
derive the RG flow starting from the Toner-Tu
equations~\cite{toner_reanalysis_2012} using the non-perturbative RG
formalism~\cite{dupuis2021nonperturbative}. Although they do not write
explicitly the dynamics of an angular variable, they find that, at the
fixed point, the scaling relation associated to the symmetry of
Eq.~(\ref{eq:NG-theta}) is obeyed exactly. Chen et
al.~\cite{chen2025response} assert that the neglected nonlinearities
may break the scaling relation but they do not support this claim,
which directly contradicts~\cite{jentsch_new_2024}, by any argument.

In contrast, we have given in~\cite{chate_dynamic_2024} an example of
the problems that arise when accidentally breaking the symmetry of
Eq.~(\ref{eq:NG-theta}) during the coarse-graining. Indeed, we showed
that writing, from the Toner-Tu equation, an equation for the phase of
the velocity containing only the relevant nonlinearities (see
Eq.(22-23) in the SM of~\cite{chate_dynamic_2024} or equivalently
Eq.(2.18,2.28) of~\cite{toner_reanalysis_2012}), one finds that this
quantity becomes massive and thus cannot describe a Goldstone
mode. This is not ``repaired'' by the RG flow which, dominated by the
mass term, instead goes towards a Gaussian fixed point which has
short-ranged correlations, and, hence, none of the scale-free behavior
of flocks. The proposal of Chen et al.~\cite{chen2025response} to turn
a blind eye and keep using these equations as a starting point of the
RG flow is thus unsatisfactory. These problems are easily fixed by
considering, instead of the phase of the velocity, a Goldstone mode
that is corrected by density fluctuations to obey
Eq.~(\ref{eq:NG-theta})~\cite{chate_dynamic_2024}.

To conclude, we have emphasized here that, in addition to the
numerical
evidence~\cite{mahault_quantitative_2019,chate_dynamic_2024}, the
symmetry Eq.~(\ref{eq:NG-theta}) for Goldstone modes associated to
rotational symmetry breaking is also supported by strong analytical
arguments.

\bibliography{refs.bib}

\begin{thebibliography}{13}%
\makeatletter
\providecommand \@ifxundefined [1]{%
 \@ifx{#1\undefined}
}%
\providecommand \@ifnum [1]{%
 \ifnum #1\expandafter \@firstoftwo
 \else \expandafter \@secondoftwo
 \fi
}%
\providecommand \@ifx [1]{%
 \ifx #1\expandafter \@firstoftwo
 \else \expandafter \@secondoftwo
 \fi
}%
\providecommand \natexlab [1]{#1}%
\providecommand \enquote  [1]{``#1''}%
\providecommand \bibnamefont  [1]{#1}%
\providecommand \bibfnamefont [1]{#1}%
\providecommand \citenamefont [1]{#1}%
\providecommand \href@noop [0]{\@secondoftwo}%
\providecommand \href [0]{\begingroup \@sanitize@url \@href}%
\providecommand \@href[1]{\@@startlink{#1}\@@href}%
\providecommand \@@href[1]{\endgroup#1\@@endlink}%
\providecommand \@sanitize@url [0]{\catcode `\\12\catcode `\$12\catcode
  `\&12\catcode `\#12\catcode `\^12\catcode `\_12\catcode `\%12\relax}%
\providecommand \@@startlink[1]{}%
\providecommand \@@endlink[0]{}%
\providecommand \url  [0]{\begingroup\@sanitize@url \@url }%
\providecommand \@url [1]{\endgroup\@href {#1}{\urlprefix }}%
\providecommand \urlprefix  [0]{URL }%
\providecommand \Eprint [0]{\href }%
\providecommand \doibase [0]{https://doi.org/}%
\providecommand \selectlanguage [0]{\@gobble}%
\providecommand \bibinfo  [0]{\@secondoftwo}%
\providecommand \bibfield  [0]{\@secondoftwo}%
\providecommand \translation [1]{[#1]}%
\providecommand \BibitemOpen [0]{}%
\providecommand \bibitemStop [0]{}%
\providecommand \bibitemNoStop [0]{.\EOS\space}%
\providecommand \EOS [0]{\spacefactor3000\relax}%
\providecommand \BibitemShut  [1]{\csname bibitem#1\endcsname}%
\let\auto@bib@innerbib\@empty
\bibitem [{\citenamefont {Chat{\'e}}\ and\ \citenamefont
  {Solon}(2025)}]{chate2025comment}%
  \BibitemOpen
  \bibfield  {author} {\bibinfo {author} {\bibfnamefont {H.}~\bibnamefont
  {Chat{\'e}}}\ and\ \bibinfo {author} {\bibfnamefont {A.}~\bibnamefont
  {Solon}},\ }\bibfield  {title} {\bibinfo {title} {Comment on the inconvenient
  truth about flocks by chen et al},\ }\href@noop {} {\bibfield  {journal}
  {\bibinfo  {journal} {arXiv preprint arXiv:2504.13683}\ } (\bibinfo {year}
  {2025})}\BibitemShut {NoStop}%
\bibitem [{\citenamefont {{Chen}}\ \emph {et~al.}(2025)\citenamefont {{Chen}},
  \citenamefont {{Jentsch}}, \citenamefont {{Lee}}, \citenamefont {{Maitra}},
  \citenamefont {{Ramaswamy}},\ and\ \citenamefont
  {{Toner}}}]{maitra_inconvenient_2025}%
  \BibitemOpen
  \bibfield  {author} {\bibinfo {author} {\bibfnamefont {L.}~\bibnamefont
  {{Chen}}}, \bibinfo {author} {\bibfnamefont {P.}~\bibnamefont {{Jentsch}}},
  \bibinfo {author} {\bibfnamefont {C.~F.}\ \bibnamefont {{Lee}}}, \bibinfo
  {author} {\bibfnamefont {A.}~\bibnamefont {{Maitra}}}, \bibinfo {author}
  {\bibfnamefont {S.}~\bibnamefont {{Ramaswamy}}},\ and\ \bibinfo {author}
  {\bibfnamefont {J.}~\bibnamefont {{Toner}}},\ }\bibfield  {title} {\bibinfo
  {title} {{The inconvenient truth about flocks}},\ }\bibfield  {journal}
  {\bibinfo  {journal} {arXiv e-prints}\ }\href
  {https://doi.org/10.48550/arXiv.2503.17064} {10.48550/arXiv.2503.17064}
  (\bibinfo {year} {2025})\BibitemShut {NoStop}%
\bibitem [{\citenamefont {Chaté}\ and\ \citenamefont
  {Solon}(2024)}]{chate_dynamic_2024}%
  \BibitemOpen
  \bibfield  {author} {\bibinfo {author} {\bibfnamefont {H.}~\bibnamefont
  {Chaté}}\ and\ \bibinfo {author} {\bibfnamefont {A.}~\bibnamefont {Solon}},\
  }\bibfield  {title} {\bibinfo {title} {Dynamic scaling of two-dimensional
  polar flocks},\ }\href {https://doi.org/10.1103/PhysRevLett.132.268302}
  {\bibfield  {journal} {\bibinfo  {journal} {Physical Review Letters}\
  }\textbf {\bibinfo {volume} {132}},\ \bibinfo {pages} {268302} (\bibinfo
  {year} {2024})}\BibitemShut {NoStop}%
\bibitem [{\citenamefont {Chen}\ \emph {et~al.}(2025)\citenamefont {Chen},
  \citenamefont {Jentsch}, \citenamefont {Lee}, \citenamefont {Maitra},
  \citenamefont {Ramaswamy},\ and\ \citenamefont {Toner}}]{chen2025response}%
  \BibitemOpen
  \bibfield  {author} {\bibinfo {author} {\bibfnamefont {L.}~\bibnamefont
  {Chen}}, \bibinfo {author} {\bibfnamefont {P.}~\bibnamefont {Jentsch}},
  \bibinfo {author} {\bibfnamefont {C.~F.}\ \bibnamefont {Lee}}, \bibinfo
  {author} {\bibfnamefont {A.}~\bibnamefont {Maitra}}, \bibinfo {author}
  {\bibfnamefont {S.}~\bibnamefont {Ramaswamy}},\ and\ \bibinfo {author}
  {\bibfnamefont {J.}~\bibnamefont {Toner}},\ }\bibfield  {title} {\bibinfo
  {title} {Response to the comment on the inconvenient truth about flocks by
  {Chat\'e} and {Solon}},\ }\href@noop {} {\bibfield  {journal} {\bibinfo
  {journal} {arXiv preprint arXiv:2505.21602}\ } (\bibinfo {year}
  {2025})}\BibitemShut {NoStop}%
\bibitem [{\citenamefont {Daviet}\ \emph {et~al.}(2024)\citenamefont {Daviet},
  \citenamefont {Zelle}, \citenamefont {Asadollahi},\ and\ \citenamefont
  {Diehl}}]{daviet2024kardar}%
  \BibitemOpen
  \bibfield  {author} {\bibinfo {author} {\bibfnamefont {R.}~\bibnamefont
  {Daviet}}, \bibinfo {author} {\bibfnamefont {C.~P.}\ \bibnamefont {Zelle}},
  \bibinfo {author} {\bibfnamefont {A.}~\bibnamefont {Asadollahi}},\ and\
  \bibinfo {author} {\bibfnamefont {S.}~\bibnamefont {Diehl}},\ }\bibfield
  {title} {\bibinfo {title} {Kardar-parisi-zhang scaling in time-crystalline
  matter},\ }\href@noop {} {\bibfield  {journal} {\bibinfo  {journal} {arXiv
  preprint arXiv:2412.09677}\ } (\bibinfo {year} {2024})}\BibitemShut {NoStop}%
\bibitem [{Note1()}]{Note1}%
  \BibitemOpen
  \bibinfo {note} {As is well known~\cite {chaikin1995principles}, there is
  then no additional Goldstone mode associated with rotational symmetry
  breaking.}\BibitemShut {Stop}%
\bibitem [{\citenamefont {Martin}\ \emph {et~al.}(1972)\citenamefont {Martin},
  \citenamefont {Parodi},\ and\ \citenamefont {Pershan}}]{martin_unified_1972}%
  \BibitemOpen
  \bibfield  {author} {\bibinfo {author} {\bibfnamefont {P.~C.}\ \bibnamefont
  {Martin}}, \bibinfo {author} {\bibfnamefont {O.}~\bibnamefont {Parodi}},\
  and\ \bibinfo {author} {\bibfnamefont {P.~S.}\ \bibnamefont {Pershan}},\
  }\bibfield  {title} {\bibinfo {title} {Unified hydrodynamic theory for
  crystals, liquid crystals, and normal fluids},\ }\href
  {https://doi.org/10.1103/PhysRevA.6.2401} {\bibfield  {journal} {\bibinfo
  {journal} {Physical Review A}\ }\textbf {\bibinfo {volume} {6}},\ \bibinfo
  {pages} {2401} (\bibinfo {year} {1972})}\BibitemShut {NoStop}%
\bibitem [{\citenamefont {Nelson}\ and\ \citenamefont
  {Pelcovits}(1977)}]{nelson_momentum-shell_1977}%
  \BibitemOpen
  \bibfield  {author} {\bibinfo {author} {\bibfnamefont {D.~R.}\ \bibnamefont
  {Nelson}}\ and\ \bibinfo {author} {\bibfnamefont {R.~A.}\ \bibnamefont
  {Pelcovits}},\ }\bibfield  {title} {\bibinfo {title} {Momentum-shell
  recursion relations, anisotropic spins, and liquid crystals in 2+epsilon
  dimensions},\ }\href@noop {} {\bibfield  {journal} {\bibinfo  {journal}
  {Physical Review B}\ }\textbf {\bibinfo {volume} {16}},\ \bibinfo {pages}
  {2191} (\bibinfo {year} {1977})},\ \bibinfo {note} {publisher:
  APS}\BibitemShut {NoStop}%
\bibitem [{\citenamefont {Jentsch}\ and\ \citenamefont
  {Lee}(2024)}]{jentsch_new_2024}%
  \BibitemOpen
  \bibfield  {author} {\bibinfo {author} {\bibfnamefont {P.}~\bibnamefont
  {Jentsch}}\ and\ \bibinfo {author} {\bibfnamefont {C.~F.}\ \bibnamefont
  {Lee}},\ }\bibfield  {title} {\bibinfo {title} {New {Universality} {Class}
  {Describes} {Vicsek}’s {Flocking} {Phase} in {Physical} {Dimensions}},\
  }\href {https://doi.org/10.1103/PhysRevLett.133.128301} {\bibfield  {journal}
  {\bibinfo  {journal} {Physical Review Letters}\ }\textbf {\bibinfo {volume}
  {133}},\ \bibinfo {pages} {128301} (\bibinfo {year} {2024})}\BibitemShut
  {NoStop}%
\bibitem [{\citenamefont {Toner}(2012)}]{toner_reanalysis_2012}%
  \BibitemOpen
  \bibfield  {author} {\bibinfo {author} {\bibfnamefont {J.}~\bibnamefont
  {Toner}},\ }\bibfield  {title} {\bibinfo {title} {Reanalysis of the
  hydrodynamic theory of fluid, polar-ordered flocks},\ }\href
  {https://doi.org/10.1103/PhysRevE.86.031918} {\bibfield  {journal} {\bibinfo
  {journal} {Physical Review E}\ }\textbf {\bibinfo {volume} {86}},\ \bibinfo
  {pages} {031918} (\bibinfo {year} {2012})}\BibitemShut {NoStop}%
\bibitem [{\citenamefont {Dupuis}\ \emph {et~al.}(2021)\citenamefont {Dupuis},
  \citenamefont {Canet}, \citenamefont {Eichhorn}, \citenamefont {Metzner},
  \citenamefont {Pawlowski}, \citenamefont {Tissier},\ and\ \citenamefont
  {Wschebor}}]{dupuis2021nonperturbative}%
  \BibitemOpen
  \bibfield  {author} {\bibinfo {author} {\bibfnamefont {N.}~\bibnamefont
  {Dupuis}}, \bibinfo {author} {\bibfnamefont {L.}~\bibnamefont {Canet}},
  \bibinfo {author} {\bibfnamefont {A.}~\bibnamefont {Eichhorn}}, \bibinfo
  {author} {\bibfnamefont {W.}~\bibnamefont {Metzner}}, \bibinfo {author}
  {\bibfnamefont {J.~M.}\ \bibnamefont {Pawlowski}}, \bibinfo {author}
  {\bibfnamefont {M.}~\bibnamefont {Tissier}},\ and\ \bibinfo {author}
  {\bibfnamefont {N.}~\bibnamefont {Wschebor}},\ }\bibfield  {title} {\bibinfo
  {title} {The nonperturbative functional renormalization group and its
  applications},\ }\href@noop {} {\bibfield  {journal} {\bibinfo  {journal}
  {Physics Reports}\ }\textbf {\bibinfo {volume} {910}},\ \bibinfo {pages} {1}
  (\bibinfo {year} {2021})}\BibitemShut {NoStop}%
\bibitem [{\citenamefont {Mahault}\ \emph {et~al.}(2019)\citenamefont
  {Mahault}, \citenamefont {Ginelli},\ and\ \citenamefont
  {Chaté}}]{mahault_quantitative_2019}%
  \BibitemOpen
  \bibfield  {author} {\bibinfo {author} {\bibfnamefont {B.}~\bibnamefont
  {Mahault}}, \bibinfo {author} {\bibfnamefont {F.}~\bibnamefont {Ginelli}},\
  and\ \bibinfo {author} {\bibfnamefont {H.}~\bibnamefont {Chaté}},\
  }\bibfield  {title} {\bibinfo {title} {Quantitative assessment of the {Toner}
  and {Tu} theory of polar flocks},\ }\href
  {https://doi.org/10.1103/PhysRevLett.123.218001} {\bibfield  {journal}
  {\bibinfo  {journal} {Physical Review Letters}\ }\textbf {\bibinfo {volume}
  {123}},\ \bibinfo {pages} {218001} (\bibinfo {year} {2019})}\BibitemShut
  {NoStop}%
\bibitem [{\citenamefont {Chaikin}\ \emph {et~al.}(1995)\citenamefont
  {Chaikin}, \citenamefont {Lubensky},\ and\ \citenamefont
  {Witten}}]{chaikin1995principles}%
  \BibitemOpen
  \bibfield  {author} {\bibinfo {author} {\bibfnamefont {P.~M.}\ \bibnamefont
  {Chaikin}}, \bibinfo {author} {\bibfnamefont {T.~C.}\ \bibnamefont
  {Lubensky}},\ and\ \bibinfo {author} {\bibfnamefont {T.~A.}\ \bibnamefont
  {Witten}},\ }\href@noop {} {\emph {\bibinfo {title} {Principles of condensed
  matter physics}}},\ Vol.~\bibinfo {volume} {10}\ (\bibinfo  {publisher}
  {Cambridge university press Cambridge},\ \bibinfo {year} {1995})\BibitemShut
  {NoStop}%
\end{thebibliography}%

\end{document}